\documentclass{emulateapj}
\usepackage{amssymb, latexsym}
\usepackage{graphicx}
\usepackage{natbib}

\begin{document}

\title{SpecPro: An Interactive IDL Program for Viewing and Analyzing Astronomical Spectra}

\author{D. Masters\altaffilmark{1,2}, P. Capak\altaffilmark{2,3}}

\altaffiltext{1}{Department of Physics and Astronomy, University of California, Riverside, CA, 92521}
\altaffiltext{2}{California Institute of Technology, 1201 East California Boulevard, Pasadena, CA 91125}
\altaffiltext{3}{Spitzer Science Center, 314-6 Caltech, Pasadena, CA, 91125}

\begin{abstract}

We present an interactive IDL program for viewing and analyzing astronomical spectra in the context of modern imaging surveys.  SpecPro's interactive design lets the user simultaneously view spectroscopic, photometric, and imaging data, allowing for rapid object classification and redshift determination.  The spectroscopic redshift can be determined with automated cross-correlation against a variety of spectral templates or by overlaying common emission and absorption features on the 1-D and 2-D spectra. Stamp images as well as the spectral energy distribution (SED) of a source can be displayed with the interface, with the positions of prominent photometric features indicated on the SED plot. Results can be saved to file from within the interface.  In this paper we discuss key program features and provide an overview of the required data formats.

\end{abstract}
\maketitle

\section{Introduction}\label{S:intro}

Modern imaging surveys such as the Cosmic Evolution Survey (COSMOS) \citep{Scoville07}) and the Great Observatories Origins Deep Survey (GOODS)  \citep{Giavalisco04} contain a wealth of galaxy information in the form of stamp images and multiwavelength photometry across the electromagnetic spectrum. Spectroscopy adds to preexisting survey data through, among other information, precise redshift determination and firm galaxy classification based on spectral features. While spectral analysis can be automated in some cases, human examination is often necessary. This is especially true when the spectroscopic targets are faint or unusual sources for which automated redshift determination is not reliable. 

Imaging and photometric data from a multiwavelength survey can play a vital role in deciphering difficult spectra. For example, a high-redshift galaxy is expected to have a strong photometric break (the ``Lyman break'') at rest-frame $\lambda$1216~\AA, with essentially no flux short of the rest-frame $\lambda$912~\AA\ Lyman limit. The presence of this spectral break in the SED can suggest an initial redshift guess and validate a high-redshift assignment based on weak spectral features. Conversely, photometric detections blueward of the Lyman limit can preclude a high-redshift assignment.  As another example, passive galaxies typically show a peak in the SED near rest-frame 1.6~$\mu$m and a spectral break near rest-frame $\lambda$4000~\AA, while their spectra reveal little more than faint absorption features. Again, the photometric characteristics can be used to inform the redshift assignment. Stamp images of a source can also be very helpful in resolving puzzling spectra, as they reveal the morphology and spatial extent of a source as well as possible sources of photometric or spectroscopic contamination. High-resolution imaging can thus help to distinguish stars from galaxies and reveal nearby or coincident systems giving rise to anamolous spectra.

With the large numbers of spectra taken in survey fields, it is necessary to have a means of rapid spectral analysis. SpecPro was developed to analyze faint spectra of high-redshift galaxies and active galactic nuclei (AGN) in the COSMOS field, where the available photometric data effectively provides low-resolution spectra covering 0.1 to 8~$\mu$m. The spectra, taken with the Keck II DEIMOS multislit spectrometer \citep{Faber03}, often show few strong features. By incorporating the SED and postage stamp images from COSMOS in the SpecPro interface we have significantly improved our ability to find redshifts and understand our galaxy sample. In addition to facilitating the analysis of faint sources, the cross-correlation capability makes redshift determination for bright sources extremely fast.  

Since SpecPro was originally designed to examine spectra taken with a multislit spectrometer, data structures are organized around the idea of a slit mask.  Data directories contain files for sources on a mask, the files within the mask directory being differentiated by slit number.  Aside from this general structure, we have made the input formats quite flexible in order to accommodate data from other instruments/surveys. In addition to DEIMOS, the code has to date been used successfully with data from the IMACS multislit spectrometer on Magellan \citep{Dressler06} as well as the FMOS fiber spectrometer and FOCAS multisplit spectrometer, both on Subaru \citep{Kashikawa02, Kimura10}. Any spectroscopic data can be viewed with SpecPro after conforming to the simple formats outlined in Section 3.

The structure of this paper is as follows. In Section 2 we provide an overview of the interface and discuss the motivation behind its design. In Section 3 we discuss the inputs to the program. Section 4 concludes with a summary and a brief discussion of possible further development. 

The code with installation instructions as well as example data, an overview of data formats, and a brief tutorial are available at http://specpro.caltech.edu. \newpage

\section{SpecPro Overview}

\begin{figure*}[ht]
        \centering
	\includegraphics[scale=0.36]{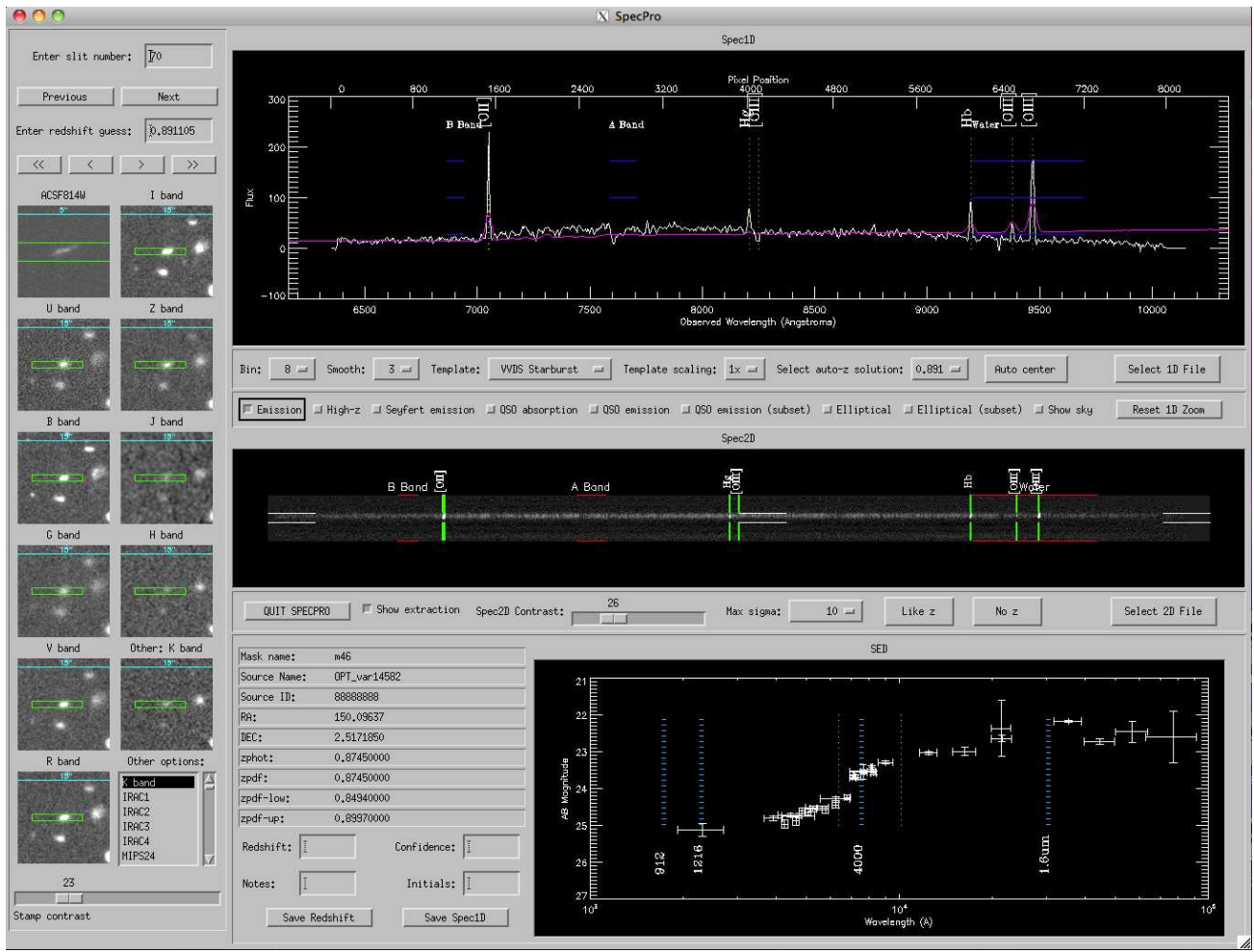}
	\caption{The SpecPro widget interface. The object shown is a star-forming galaxy at redshift 0.89. Stamp images of the source are displayed on the left side of the interface, and the 1-D spectrum is displayed in the upper right window. On the 1-D plot, vertical lines mark the positions of prominent emission features, and a spectral template is overlaid in blue. Under the 1-D spectrum is the corresponding 2-D spectrum, binned to fit the display. The same emission lines plotted over the 1-D spectrum are indicated in the 2-D spectrum, and the atmospheric absorption regions are marked with horizontal red lines above and below the spectrum. Emission lines ([OII], H$\gamma$, H$\beta$, [OIII]) are easily visible in the binned 2-D image. The window on the bottom right shows the SED of the source from the ultraviolet to the mid-infrared. The AB magnitude (increasing downwards) is plotted against filter wavelength. The SED data points indicate filter FWHM with horizontal bars and photometric errors with vertical bars. The observed-frame positions of important photometric indicators, including the Lyman limit at  $\lambda$912~\AA, Ly$\alpha$ at $\lambda$1216~\AA, the 4000~\AA\ break, and 1.6~$\mu$m are indicated with vertical dashed lines.}
\label{Fi:plotone}
\end{figure*}

\begin{figure*}[ht]
        \centering
	\includegraphics[scale=0.36]{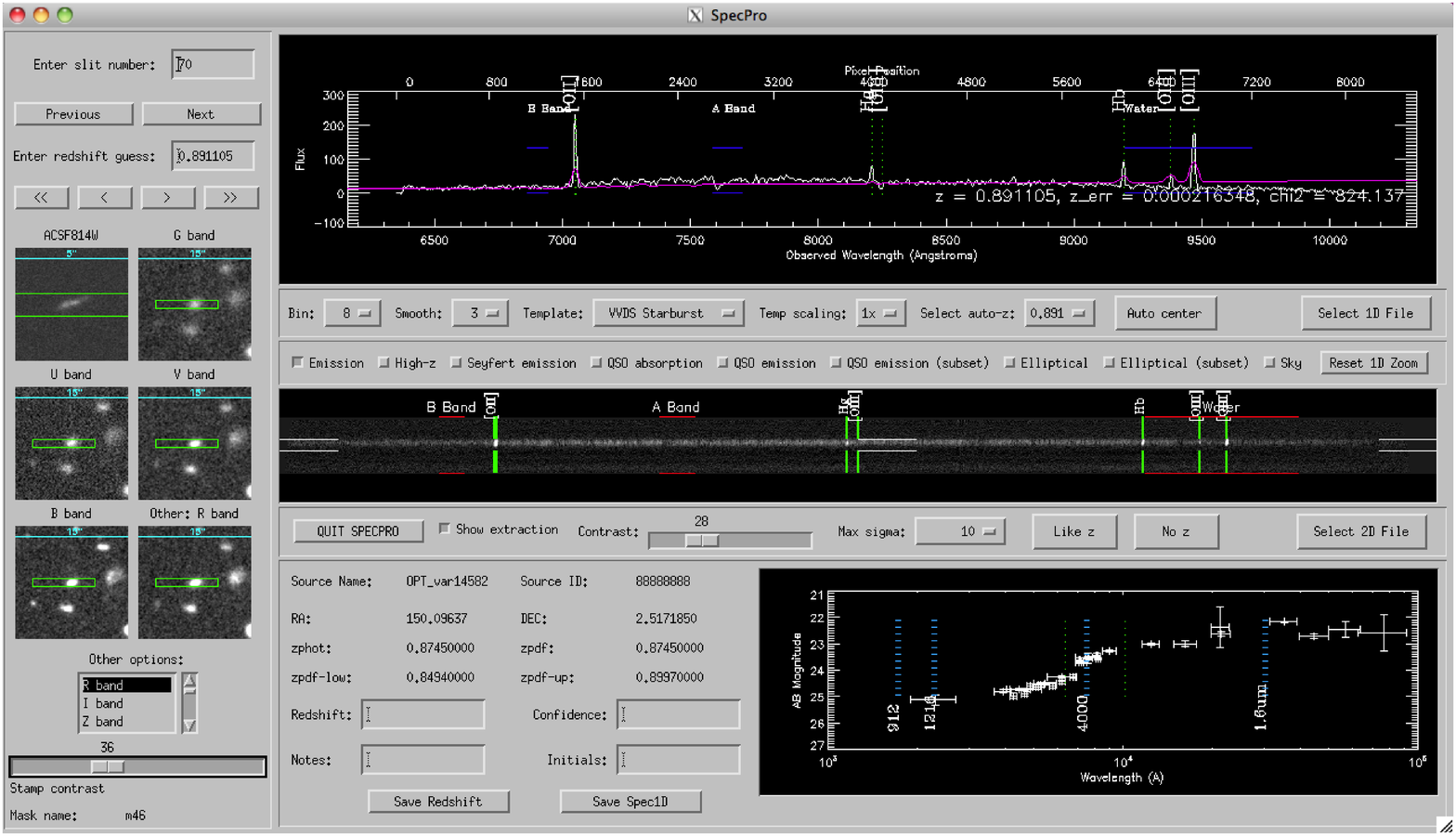}
	\caption{The SpecPro widget interface, small version.}
\label{Fi:plottwo}
\end{figure*}

The SpecPro interface is shown in Figures~\ref{Fi:plotone} and \ref{Fi:plottwo}. Figure~\ref{Fi:plotone} shows the full interface, while Figure~\ref{Fi:plottwo} shows a smaller version that can be invoked for use on smaller monitors. A galaxy with strong emission lines from the COSMOS survey is displayed.

Surrounding the data displays are widgets allowing the user to perform various tasks, such as navigate through spectra in the mask directory, adjust the current redshift guess, bin and smooth the 1-D spectrum, overlay galaxy templates, plot the redshifted positions of common emission and absorption features, change stamp image and 2-D spectrum contrast, and save results to an external file. In the following sections we outline the key capabilities of the interface, emphasizing how they assist in spectral analysis.

It is worth emphasizing that the SpecPro interface does not require all of the various pieces of information (2-D spectrum, stamp images, SED) to run. It can be used with a subset of this information and the windows for which there is no data are left blank. Therefore the interface can be used to view and analyze spectra for which the data associated with a multiwavelength survey is absent, or to pre-select targets for spectroscopy using just photometric and imaging data.

\subsection{Stamp images}

Available stamp images of the source are displayed on the left side of the interface in 100x100 pixel windows. Under the final window is a scroll list containing the names of overflow images.  The user can click entries in this list to display them in the final stamp window, allowing an arbitrary number of stamp images to be stored and viewed with the interface. The images are displayed in the order in which they are stored in the structure array, beginning with the upper left stamp window and going down.

The displayed images are signal-to-noise (S/N) plots generated by dividing the image flux by the average error in the image. This maximizes the dynamic range in order to display both bright sources and faint counterparts blueward of a spectral break. By viewing the stamps the user can quickly determine the source of a serendipitous spectrum and whether photometric points in the source SED correspond to real detections or artifacts. 

Selecting the ``Show extraction" button (located underneath the 2-D plot window) causes the scale in arcseconds to be displayed on the stamps. In addition, if the slit information (RA, DEC, length, width, and angle) is provided, it is plotted over the stamp images, as in Figures~\ref{Fi:plotone} and \ref{Fi:plottwo}. 

\subsection{Spectrum Display}

The one-dimensional spectrum is displayed in the upper right window of the display. The spectrum is shown in white, with overlaid spectral templates shown in blue. Regions of significant atmospheric absorption are indicated with horizontal blue lines. The residual error, which can help determine whether a weak feature is real or an artifact due to sky subtraction, can be overlaid on the plot by clicking the ``Show sky'' button on the second row down from the 1-D display. 

The 1-D spectrum can be binned and smoothed to bring out features in low S/N spectra. Binning combines pixels, reducing the resolution of the spectrum to increase the S/N, while smoothing performs an inverse variance weighted average in the vicinity of each pixel to reduce the noise. In Figure~\ref{Fi:plotone} the spectrum has been binned and smoothed, clearly revealing prominent emission features. 

The 2-D spectrum is displayed beneath the 1-D spectrum, binned, if necessary, to fit the display. The 2-D spectrum can be helpful to: (1)~verify the fidelity of the 1-D spectrum, (2)~determine whether features observed in the 1-D spectrum are real or artifacts, (3)~identify serendipitous objects, and (4)~examine the spatial morphology of emission features. As described in Section 2.3, the user has the ability to reextract the 1-D spectrum from the 2-D, which can be very useful in cases in which there was a problem with 1-D extraction or the 1-D spectrum of a serendipitous source is desired.

The contrast of the 2-D spectrum can be adjusted with a slider widget. Next to the contrast slider is a droplist widget labeled ``Max sigma:". This is set to 10 by default, which indicates that contrast scaling has an upper limit of 10$\sigma$, so that pixels 10$\sigma$ from the mean are rescaled to the maximum pixel value of 255. Lowering this value can help in cases in which the spectrum of interest is being drowned out by the light of bright serendipitous sources.

The positions (observed-frame) of emission and absorption features of different galaxy types can be displayed over both the 1-D and the 2-D spectrum by selecting the buttons above the 2-D display. Typical lines from star-forming galaxies, passive galaxies, quasistellar objects (QSOs), Seyferts, and high-redshift galaxies can be selected.  Redshifts can be found manually by overlaying the appropriate lines on the spectrum and adjusting the redshift until they line up with observed spectral features. These lines can also be used to identify absorption and emission features in the various spectral templates.

Both the 1-D and the 2-D spectrum can be zoomed on by clicking, dragging and releasing around the region of interest. In the case of the 2-D, this action causes the zoomed region to be displayed in its unbinned state in a separate window. This can be helpful, for instance, to determine whether an emission feature is [OII] or Ly$\alpha$, which can often be deduced from the morphology of the emission line. 

\subsection{Reextraction of 1-D spectrum}
It is often necessary to reextract the 1-D spectrum from the 2-D, particularly in order to recover serendipitously observed sources that fall in the slit. We have provided a method of easily reextracting the 1-D spectrum from the 2-D, adapting the routine extract1d.pro from the DEIMOS DEEP2 pipeline code \citep{Marinoni01} for this purpose. To reextract, the user places the cursor over the 2-D spectrum at the vertical position desired, then clicks, drags horizontally (by any amount, in either direction) and releases. This triggers the reextraction, and when it is complete the 1-D plot is updated with the newly extracted spectrum. A dialog box lets the user save the extracted 1-D spectrum in .fits format.

\subsection{SED}
The source SED is displayed in the bottom right window of the interface. The information for this plot comes from the photometry file, described in Section 3.5. The plot extends from 0.1~to~10~$\mu$m and displays photometric AB magnitudes with errors. The effective widths of the photometric bands are indicated with horizontal bars and errors in the magnitude estimates are indicated with vertical bars. Observed-frame positions of important SED features are indicated with vertical blue lines. These features include the Lyman limit at 912~\AA, Ly$\alpha$ at 1216~\AA, the 4000~\AA~break, and the typical SED peak of evolved galaxies at 1.6~$\mu$m.  The wavelength coverage of the spectrum is indicated with vertical green lines.  

\subsection{Cross-Correlation}
Automated redshift determination by convolution against spectral templates is a powerful and time-saving capability. We have adapted cross-correlation routines originally written for the SDSS spectral reduction package for use in SpecPro, with a library of spectral templates available to correlate against (see Table~1). The templates span a broad range of galaxy types. We have also included some stellar templates to help identify high-redshift selected sources that are actually stars. 

The user is expected to be able to choose an appropriate template based on the spectrum and other information about the source. When one of the galaxy templates from the template library is selected, the six best redshift matches are computed and the best solution is displayed. The droplist ``Auto-z solution:" under the 1-D spectrum contains all of the solutions for examination. Clicking an alternate solution will switch to the new redshift guess and move the template accordingly. Because the spectral template is overplotted on the 1-D spectrum, the accuracy of the result can be easily checked. The template can be scaled via the ``Template scaling:'' button to provide a better visual match to the spectrum.

In some cases cross-correlation fails to find the solution, even though the correct answer is apparent to the user.  In this case the redshift can be adjusted manually to find the answer, but finding the exact solution can be time consuming. Therefore we added an ``Auto center" button that cross-correlates against the currently selected galaxy template in a small region (z~$\pm$~0.05) around the current redshift guess. 

In some very difficult cases cross-correlation will fail to find the correct redshift, even using auto-centering. In these unusual instances the user must resort to adjusting the redshift manually to the correct answer.

\begin{deluxetable*}{ll}
\tabletypesize{\scriptsize}
 \tablecaption{Summary of the spectral templates available for cross-correlation in SpecPro.}\label{Ta:templates}
 \tablehead{\colhead{\bf Template} & \colhead{\bf Description}} 
  \startdata
      VVDS LBG & Lyman-break galaxy template from the VIRMOS-VLT Deep Survey \citep{Lefevre05} \\ 
      VVDS Elliptical & Passive galaxy template, strong absorption, no emission \\ 
      VVDS S0 &  S0 classification template, weak emission, absorption \\ 
      VVDS Early Spiral &  Stronger emission, less absorption \\ 
      VVDS Spiral &  Typical spiral spectrum, with strong emission features \\ 
      VVDS Starburst &  Extremely strong nebular emission features \\ 
      SDSS Quasar & Typical broad-line quasar spectrum from the Sloan Digital Sky Survey \citep{Schneider10} \\ 
      Red Galaxy &  Passive galaxy template from PEGASE spectral evolution model \citep{Fioc97} \\
      Green Galaxy & Early spiral/spiral template from PEGASE \\ 
      Blue Galaxy & Spiral/Starburst template from PEGASE \\ 
      LBG Shapley & Lyman-break galaxy template from Shapley 2003 \citep{Shapley03} \\ 
      SDSS LoBAL & Low-ionization broad absorption line (BAL) quasar template \citep{Reichard03} \\ 
      SDSS HiBAL & High-ionization BAL quasar template \citep{Reichard03} \\ 
      A0-M6 & Stellar templates from the Pickles catalog \citep{Pickles98} \\ 
      \enddata
\end{deluxetable*}

\subsection{Output}
The interface allows the user to rapidly record results to a formatted file. To the left of the SED plot window are four output fields with which the user can save the redshift, a confidence associated with it, their initials, and notes on the spectrum. The output is saved in ascii format and contains the mask name, source name, slit number, RA, and DEC as well as the redshift result. Results are always appended to the end of the selected output file and multiple entries for a single source can be added. Once an output file has been selected, it is the default output location for the rest of the SpecPro session. 

The ``Save Spec1D" button saves the displayed 1-D spectrum (wavelength and flux at the current zoom, bin, and smooth level) to an ascii file for plotting in other programs. In addition, images in .tif format of any of the data windows can be saved by double-clicking on them. 

\section{Data Formats}

We have attempted to make the required formats simple and general so that data from any instrument/survey can be easily adapted for use with SpecPro.  While the files are expected to be arranged by slit number, any sequential numbering scheme can be used to differentiate sources. None of the inputs accepted by SpecPro are actually required for the program to run; if one is missing, the corresponding field in the interface is left blank. The program accepts the following five files$\colon$~\begin{itemize}
\item 1-D spectrum file (fits)
\item 2-D spectrum file (fits)
\item Stamp image file (fits)
\item Photometry file (ascii)
\item Information file (ascii)
\end{itemize}

These are briefly described in the rest of this section. More detailed descriptions are given on the website, and example data is provided with the program download.

\subsection{1-D Spectrum}

The 1-D spectrum is a stored as a structure with three fields$\colon$ \emph{flux}, \emph{ivar}, and \emph{lambda}. Each field holds a one-dimensional double array. The \emph{flux} field is the spectrum, the \emph{ivar} field is the inverse variance of the flux, and \emph{lambda} is the wavelength at each pixel, in angstroms. Note that the units of flux are arbitrary, but values proportional to F$_\nu$ are preferable because this matches the units of the spectral templates. The structure must be saved in FITS format.

\subsection{2-D Spectrum}

The 2-D spectrum is also stored in a structure with the three fields \emph{flux}, \emph{ivar}, and \emph{lambda}. However, each field here is a two-dimensional double array, including the \emph{lambda} field, which contains the wavelength solution for each pixel in the 2-D spectrum. The 2-D spectrum is stored in FITS format.

\subsection{Stamp Images}
 
Stamp images of the source are stored in a single FITS file, with the data stored as an array of structures. Each element in the array contains a stamp image and ancillary information, with the size of the array determined by how many stamps are available. The stamp structures must all be of the same form, with fields as follows$\colon$

\begin{itemize}
\item \emph{name} -  string giving name of the image, e.g. ``U band"
\item \emph{flux} - 100x100 double array
\item \emph{ivar} - 100x100 double array
\item \emph{RA} - image center in degrees, double
\item \emph{DEC} - image center in degrees, double
\item \emph{pixscale} - arcseconds per pixel, double
\end{itemize}

The stamps should be oriented with north up. If the inverse variance for the image is not available then it should either be approximated in some way or set to all zeros. 

\subsection{Information File}

The information file contains ancillary information about a source displayed by the interface and/or used in some of its functionality. This information includes source RA and DEC, slit information, photometric redshift estimate, and other quantities (see the website for a detailed description). If the user only has a subset of the information they can omit the missing values, or set them to zero.

\subsection{Photometry}

The photometry file is a space delimited file with five columns: filter name, filter central wavelength, effective filter bandwidth, AB magnitude, and AB magnitude error. Both the filter central wavelength and effective filter bandwith should be expressed in angstroms.  Non-detections are indicated with an AB magnitude of -99, with the AB magnitude error set to the lower bound on the AB magnitude established by the non-detection. Any photometric point can be included, but the program's SED plot extends from 0.1 to 10.0~$\mu$m.

\section{Summary}

SpecPro is an interactive IDL-based interface particularly suited to viewing astronomical spectra in the context of multiwavelength surveys. In addition to 1-D and 2-D spectra, the interface can display stamp images and the spectral energy distribution of the source. Having all of this information in one display makes it much easier to identify the nature and redshift of faint spectral targets. 

We have made an effort to make the data formats required by the program both simple and general. The program could potentially incorporate wrappers for various instrument's data or allow more flexible inputs; however, the effort required to achieve this flexibility is probably too high considering the relative ease with which data can be made to conform with the formats outlined in Section 3.

One possibility for future development is to incorporate tools for more detailed spectroscopic analysis (line-fitting, determination of equivalent widths, etc.) from within the interface. This was not the our intention when designing SpecPro, but the addition of such functionality would make it a much more general tool for spectral analysis. Another related possibility would be to allow the user to interactively fit the source SED from within the interface, in order to determine galaxy stellar mass, age, and other parameters. However, at present the interface is optimized for spectral classification and redshift determination and we have no concrete plans for further development.

We have found the program valuable in our own analysis of DEIMOS spectra of targets in the COSMOS survey field. In addition to making the identification of redshifts very fast, it has also allowed us to efficiently characterize our sample and identify particularly interesting galaxies for follow-up study. We feel that SpecPro will prove useful to others working with spectroscopic samples as well.
\acknowledgments
The authors would like to thank Dr. Yuko Kakazu, Dr. Hai Fu, Dr. Lin Yan, and Dr. Nick Scoville from the California Institute of Technology for their helpful comments and suggestions during the development of the software.  In addition, we would like to acknowledge and thank Dr. Mara Salvato of the Max Planck Institute for Plasma Physics and Dr. Francesca Civano of the Harvard-Smithsonian Center for Astronomy for providing useful feedback on early versions of the code. We also thank Dr. Bahram Mobasher of the University of California, Riverside for carefully reading a draft of this paper and making suggestions that significantly improved its content. 

SpecPro makes use of IDL code written and maintained by others. We thank Wayne Landsman at NASA/GSFC for his work maintaining the IDL Astronomy User's Library. We also thank the authors of code we have used for our implementation of automated cross-correlation, including David Schlegel, Doug Finkbeiner, Michael Cooper, and John Johnson. We thank Craig Markwardt, whose MPFIT least-squares fitting package \citep{Markwardt09} is integral to the cross-correlatation functionality.

Finally, we thank an anonymous referee for carefully reading this manuscript and providing very constructive feedback both in terms of content and with regard to the distribution of the software on the web.

This work was supported in part by a visiting graduate student fellowship at the Caltech Infrared Processing and Analysis Center (IPAC).

\bibliographystyle{plainnat}
\bibliography{specref}

\end{document}